\documentclass[24pt]{article}
\usepackage{amssymb}
\usepackage{amsfonts}
\usepackage{amssymb,amsmath}
\usepackage{mathrsfs}
\usepackage{latexsym}
\usepackage{amsmath}
\usepackage{latexsym}
\usepackage[cp1251]{inputenc}
\usepackage{graphicx}
\usepackage{color}

\title{An application of a quantum wave impedance method to finite periodic structures}
\author{O. I. Hryhorchak\\
{\small Department for Theoretical Physics, Ivan Franko National
University of Lviv,}\\
{\small 12, Drahomanov Str., Lviv, UA--79005,
Ukraine}\\
\small{\it{Orest.Hryhorchak@lnu.edu.ua}}}

\def\ch{\mathop{\rm ch}\nolimits}
\def\sh{\mathop{\rm sh}\nolimits}
\def\th{\mathop{\rm th}\nolimits}

\def\tg{\mathop{\rm tg}\nolimits}
\def\ctg{\mathop{\rm ctg}\nolimits}

\begin{document}
\renewcommand{\abstractname}{Abstract}
\maketitle

\begin{abstract}
The relations between a quantum wave impedance function  and elements  of transfer and scattering matrixes for quantum mechanical systems with arbitrary localized form of potential were established. Obtained results allows using the advantages of  both methods, namely a transfer matrix technique and a quantum wave impedance approach, for an  investigating of nanosystems with a complicated geometry of a potential. A finit Dirac comb was solved and expressions for Tamm's levels in this system were derived within both approaches.
\end{abstract}

\section{Introduction}
Heterostructures with an arbitrary form of a potential energy inside can be handled through the discretization of potential changes and following using a variety of different methods. Among the most popular are a transfer matrix approach \cite{Ando_Itoh:1987, Griffiths_Steinke:2001, Pereyra_Castillo:2002, SanchezSoto_atall:2012,Harwit_Harris_Kapitulnik:1986,Capasso_Mohammed_Cho:1986,Miller_etall:1985, Pichard:1991, Pendry_MacKinnon_Roberts:1992, Markos:2006},  a finite difference method \cite{Zhou:1993, Grossmann_Roos_Stynes:2007}, a quantum wave impedance approach \cite{Kabir_Khan_Alam:1991, Nelin_Imamov:2010, Babushkin_Nelin:2011_1, Khatyan_Gindikina_Nelin:2015, Ashby:2016}. 

Both  the transfer matrix technique and a quantum wave impedance method \cite{Mendez_Dominguez_Macia:1993, Londergan_Carini_Murdoc:1999} are widely used for solving quantum systems with a complicated geometry of a potential. These formalisms can be used for studying a propagation of both
quantum particles and electromagnetic, acoustic or elastic waves. They allow calculating a transmission and a reflection amplitudes as well as finding energies of bound states. 

Within a transfer matrix approach a wave function and its first derivative (in an arbitrary region with a constant value of potential energy) are related by a transfer $2\times2$ matrix with the ones of the next region.
The main advantage of this method is that it enables a simple representation of a solution of a Shr\"{odinger} equation through cascading $2\times2$ matrices. But this approach also has its drawbacks \cite{Arx1:2020, Arx2:2020, Arx6:2020}.
Using a quantum wave impedance method instead of two boundary conditions for a wave function and its first derivative it is  only one boundary condition for a quantum wave impedance function. It allows signifacant simplifying the process of iterative calculation of a quantum wave impedance, which we have illustarted in our previous papers \cite{Arx3:2020, Arx4:2020, Arx5:2020}.

Thus the question arise: how to combine advantages of both approaches especially in the area of investigating of finit periodic systems since a quantum wave impedance approach has much simple procedure of its iterative determination while within a transfer matrix formalizm the approach for a calculation of sequence of finit number of identical potential regions is well developed  
\cite{Griffiths_Steinke:2001, SanchezSoto_atall:2012}. Finding an answer for this question is the issue of this paper.

\section{Relation between a transfer matrix and a quantum wave impe\-dance}
Consider a localized potential of a following  form
\begin{eqnarray}\label{U1f2}
U(x)=\left\{\begin{array}{cc}
U_1, & x\le a\\
f(x),& a<x<b \\
U_2, &  x\ge b
\end{array}\right..
\end{eqnarray}
The motion of a quantum particle in the presence of the potential in 1D is governed by a Schr\"{o}dinger equation.
Its solution in the region $x\le a$ is
\begin{eqnarray}
\psi_{x\leq a}(x)=A_{+}\exp[ik_1x]+A_{-}\exp[-ik_1x]
\end{eqnarray} 
and in the region $x\ge b$:
\begin{eqnarray}
\psi_{x\geq b}(x)=B_{+}\exp[ik_2x]+B_{-}\exp[-ik_2x],
\end{eqnarray}
where $k_1=\sqrt{2m(E-U_1)}/\hbar$, $k_2=\sqrt{2m(E-U_2)}/\hbar$.

Usually,
one relates the amplitudes $A_{+}$, $A_{-}$ of the waves to the left side of the potential to those $B_{+}$, $B_{-}$ on the right side. This defines the transfer matrix $T$. 

\begin{eqnarray}
\begin{pmatrix}
A_{+}\exp[ik_1a]
\\
A_{-} \exp[-ik_1a]
\end{pmatrix}
=T_{ab}
\begin{pmatrix}
B_{+}\exp[ik_2b]
\\
B_{-}\exp[-ik_2b]
\end{pmatrix}.
\end{eqnarray}
It is well-known \cite{SanchezSoto_atall:2012} that components of a transfer matrix can be expressed through a transmission $t$ and a reflection $r$ amplitudes
:
\begin{eqnarray}\label{Tab_rt}
T_{ab}=\begin{pmatrix}
\frac{1}{t_{ab}} \quad \frac{r^*_{ab}}{t^*_{ab}}
\\
\frac{r_{ab}}{t_{ab}} \quad \frac{1}{t^*_{ab}}
\end{pmatrix}
\end{eqnarray}
and
\begin{eqnarray}
T_{ba}=\begin{pmatrix}
\frac{1}{t_{ab}} \quad -\frac{r_{ab}}{t_{ab}}
\\
-\frac{r^*_{ab}}{t^*_{ab}} \quad \frac{1}{t^*_{ab}}
\end{pmatrix},
\end{eqnarray}
where $t_{ab}$ is the transmission amplitude, $r_{ab}$ is the reflection amplitude.
Both these result and the condition $\det|T_{ab}|=1$ are
the consequence of a time-reversal invariance and a conservation of a current
density.

We see that $t_{ab}=t_{ba}$ and $\frac{r_{ba}}{t_{ba}}=-\frac{r^*_{ab}}{t^*_{ab}}$ because the scattering is symmetric with respect to the time inversion and $|t|^2+|r|^2=1$ which is the result of a conservation of a flux.

The transfer matrix approach gives us the possibility to study the properties of the sample through scattering
experiments. Assume that far from the sample we have a plane wave with an energy $E$. Then the transmission
coefficient T, the reflection coefficient R and all elements of a transfer matrix are functions of $E$. Thus, we are able to study the properties of the system on the base of its scattering
response. 

Now our task is to find the relation between a transfer matrix formalism and a quantum wave impedance approach. It will enable us to transfer the results obtained for a transfer matrix to the case of a quantum wave impedance.

It follows from the definition of a quantum wave impedance that
\begin{eqnarray}
Z(a)=z_1\frac{A_{+}e^{ik_1a}-A_{-}e^{-ik_1a}}{A_{+}e^{ik_1a}+A_{-}e^{-ik_1a}},\qquad
Z(b)=z_2\frac{B_{+}e^{ik_2a}-B_{-}e^{-ik_2a}}{B_{+}e^{ik_2a}+B_{-}e^{-ik_2a}},\nonumber
\end{eqnarray}
where $z_1=\sqrt{2(E-U_1)/m}=\frac{\hbar k_1}{m}$ and $z_2=\sqrt{2(E-U_2)/m}=\frac{\hbar k_2}{m}$ are characteristic impedances of the regions to the left and to the right of the localized potential described by a function $f(x)$ (\ref{U1f2}), $Z(a)$ and $Z(b)$ is the value of a quantum wave impedance of a studied system at points $x=a$ and $x=b$. 

From the other side the definition of a transfer matrix gives 
\begin{eqnarray}
A_{+}e^{ik_1a}=T_{11}B_{+}e^{ik_2a}+T_{12}B_{-}e^{-ik_2a},\nonumber\\
A_{-}e^{-ik_1a}=T_{21}B_{+}e^{ik_2a}+T_{22}B_{-}e^{-ik_2a},
\end{eqnarray}
Combining these definitions one gets the following relation:  
\begin{eqnarray}
Z(a)=z_1\frac{(T_{11}-T_{21})(z_2+Z(b))+(T_{12}-T_{22})(z_2-Z(b))}{(T_{11}+T_{21})(z_2+Z(b))+(T_{12}+T_{22})(z_2-Z(b))}
\end{eqnarray}
or
\begin{eqnarray}\label{Z_finit_periodic}
Z(a)\!=\!z_1\frac{z_2(T_{11}\!-\!T_{21}\!+\!T_{12}\!-\!T_{22})\!+\!Z(b)[T_{11}\!-\!T_{21}\!-\!T_{12}\!+\!T_{22}]}{z_2(T_{11}\!+\!T_{21}\!+\!T_{12}\!+\!T_{22})\!+\!Z(b)[T_{11}\!+\!T_{21}\!-\!T_{12}\!-\!T_{22}]}\!.
\end{eqnarray}
After introducing $Z_{ij}$ values ($i,j=1,2$) the last relation can be written as
\begin{eqnarray}\label{Zmatrix_finding}
Z(a)=z_1\frac{z_2Z_{11}-Z(b)Z_{12}}
{z_2Z_{21}-Z(b)Z_{22}}
\end{eqnarray}
or in notations of a \cite{Arx8:2020}:
\begin{eqnarray}
Z(a)\rightarrow z_1
\begin{pmatrix}
Z_{11} \ \ Z_{12}\\
Z_{21} \ \  Z_{22}
\end{pmatrix}
\begin{pmatrix}
z_2\\
-Z(b)
\end{pmatrix}.
\end{eqnarray} 
The relations between $Z_{ij}$ and $T_{ij}$ are as follows
\begin{eqnarray}\label{Z_thr_T}
Z_{11}&=&\frac{1}{2}\left(T_{11}-T_{21}+T_{12}-T_{22}\right),\nonumber\\
Z_{12}&=&\frac{1}{2}\left(-T_{11}+T_{21}+T_{12}-T_{22}\right),\nonumber\\
Z_{21}&=&\frac{1}{2}\left(T_{11}+T_{21}+T_{12}+T_{22}\right),\nonumber\\
Z_{22}&=&\frac{1}{2}\left(-T_{11}-T_{21}+T_{12}+T_{22}\right).
\end{eqnarray}
A multiplayer $\frac{1}{2}$ is introduced in order to $\det|Z|=\det|T|=1$. We also can express $T_{ij}$ through $Z_{ij}$, namely
\begin{eqnarray}\label{T_thr_Z}
T_{11}&=&\frac{1}{2}\left(Z_{11}-Z_{12}+Z_{21}-Z_{22}\right),\nonumber\\
T_{12}&=&\frac{1}{2}\left(Z_{11}+Z_{12}+Z_{21}+Z_{22}\right),\nonumber\\
T_{21}&=&\frac{1}{2}\left(-Z_{11}+Z_{12}+Z_{21}-Z_{22}\right),\nonumber\\
T_{22}&=&\frac{1}{2}\left(-Z_{11}-Z_{12}+Z_{21}+Z_{22}\right).
\end{eqnarray}
On the base of the result of \cite{Arx7:2020} if we have a function $f_k(b,a)$ for a considered localized potential described by a function $f(x)$ (\ref{U1f2}) then we can find both values $Z_{ij}$ and $T_{ij}$:
\begin{eqnarray}
Z_{11}=\frac{f_k^{(b,a)}(b,a)}{2z_1z_2},\quad Z_{12}=\frac{f_k^{(a)}(b,a)}{2z_2},\nonumber\\
Z_{21}=\frac{f_k^{(b)}(b,a)}{2z_1},
\quad Z_{22}=\frac{f_k(b,a)}{2},
\end{eqnarray} 

\begin{eqnarray}\label{T_thr_f}
T_{11}&=&-\frac{z_1z_2f_k(b,a)+z_2f_k^{(a)}(b,a)-z_1f_k^{(b)}(b,a)-f_k^{(b,a)}(b,a)}{2z_1z_2},\nonumber\\
T_{12}&=&\frac{z_1z_2f_k(b,a)+z_2f_k^{(a)}(b,a)+z_1f_k^{(b)}(b,a)+f_k^{(b,a)}(b,a)}{2z_1z_2},\nonumber\\
T_{21}&=&-\frac{z_1z_2f_b(b,a)-z_2f_k^{(a)}(b,a)-z_1f_k^{(b)}(b,a)+f_k^{(b,a)}(b,a)}{2z_1z_2},\nonumber\\
T_{22}&=&\frac{z_1z_2f_k(b,a)-z_2f_k^{(a)}(b,a)+z_1f_k^{(b)}(b,a)-f_k^{(b,a)}(b,a)}{2z_1z_2}.\nonumber\\
\end{eqnarray}
For a singular rectangular barrier of $U_b$ height and $L=b-a$ width assuming $E<U_b$ we have
\begin{eqnarray}
&&z_1=z_2=z_0=\sqrt{\frac{2E}{m}},\quad k_0=\frac{m}{\hbar}z_0,\quad f_b(b,a)=2\sh[\varkappa_bL],\nonumber\\
&&z_b=\sqrt{\frac{2(E-U_b)}{m}},\quad
\varkappa_b=\frac{\sqrt{2m(U_b-E)}}{\hbar}
\end{eqnarray} 
and thus
\begin{eqnarray}
Z=\begin{pmatrix}
\frac{\varkappa_b^2}{k_0^2}\sh(\varkappa_bL) & i\frac{\varkappa_b}{k_0}\ch(\varkappa_bL)
\\
-i\frac{\varkappa_b}{k_0}\ch(\varkappa_bL) & \sh(\varkappa_bL),
\end{pmatrix},
\end{eqnarray}

\begin{eqnarray}
T=\begin{pmatrix}
-i\frac{\varkappa_b^2-k_0^2}{2k_0\varkappa_b}\!\sh(\varkappa_bL)\!+\!\ch(\varkappa_bL)\!\!\!\!\!\!\!\!\! & -i\frac{k_0^2+\varkappa_b^2}{2k_0\varkappa_b}\!\sh(\varkappa_bL)
\\
i\frac{k_0^2+\varkappa_b^2}{2k_0\varkappa_b}\!\sh(\varkappa_bL)\!\!\!\!\!\!\!\!\! & -i\frac{\varkappa_b^2-k_0^2}{2k_0\varkappa_b}\!\sh(\varkappa_bL)\!+\!\ch(\varkappa_bL)
\end{pmatrix}.
\end{eqnarray}

\section{Relation between a scattering matrix and a quan\-tum wave impedance}
Although the transfer matrix approach is more widely used for an analysis of one-dimensional quantum systems the scattering matrix formalism allows formulating the physical properties of the scattering more easily. Like in a case of a transfer matrix the elements of a scattering matrix completely characterize the scattering
and transmission properties of a one-dimensional potential.
Scattering matrix describes the outgoing
waves in terms of the ingoing waves, namely
\begin{eqnarray}
\begin{pmatrix}
B_+
\\
A_- 
\end{pmatrix}=
S_{ab}
\begin{pmatrix}
A_+
\\
B_- 
\end{pmatrix},
\end{eqnarray}
It is well-known that there is a clear correspondence between components of a transfer matrix and a scattering matrix \cite{Markos_Soukoulis:2008}:
\begin{eqnarray}\label{T_thr_S}
T_{11}&=&\frac{1}{S_{11}},\qquad
T_{12}=-\frac{S_{12}}{S_{11}},\nonumber\\
T_{21}&=&\frac{S_{21}}{S_{11}},\qquad
T_{22}=S_{22}-\frac{S_{12}S_{21}}{S_{11}}
\end{eqnarray}
and
\begin{eqnarray}\label{S_thr_T}
S_{11}&=&\frac{1}{T_{11}},\qquad
S_{12}=-\frac{T_{12}}{T_{11}},\nonumber\\
S_{21}&=&\frac{T_{21}}{T_{11}},\qquad
S_{22}=T_{22}-\frac{T_{12}T_{21}}{T_{11}}.
\end{eqnarray}
Using (\ref{Tab_rt}) we easily get the components of a scattering matrix expressed through a transmission and a reflection amplitudes:
\begin{eqnarray}
S_{ab}=\begin{pmatrix}
t_{ab} \quad r_{ba}
\\
r_{ab} \quad t_{ba}
\end{pmatrix}.
\end{eqnarray}
On the base of (\ref{S_thr_T}) and (\ref{T_thr_f}) we have
\begin{eqnarray}
\!\!\!S_{11}\!\!\!&=&\!\!\!t_{ab}\!=\!-\!\frac{2z_1z_2}{z_1z_2f(b,a)\!+\!z_2f_k^{(a)}(b,a)\!-\!z_1f_k^{(b)}(b,a)\!-\!f_k^{(b,a)}(b,a)},\nonumber\\
\!\!\!S_{12}\!\!\!&=&\!\!\!r_{ba}=\frac{z_1z_2f_k(b,a)+z_2f_k^{(a)}(b,a)+z_1f_k^{(b)}(b,a)+f_k^{(b,a)}(b,a)}{z_1z_2f_k(b,a)+z_2f_k^{(a)}(b,a)-z_1f_k^{(b)}(b,a)-f_k^{(b,a)}(b,a)},\nonumber\\
\!\!\!S_{21}\!\!\!&=&\!\!\!r_{ab}=
\frac{z_1z_2f_k(b,a)-z_2f_k^{(a)}(b,a)-z_1f_k^{(b)}(b,a)+f_k^{(b,a)}(b,a)}{z_1z_2f_k(b,a)+z_2f_k^{(a)}(b,a)-z_1f_k^{(b)}(b,a)-f_k^{(b,a)}(b,a)},\nonumber\\
\!\!\!S_{22}\!\!\!&=&\!\!\!t_{ba}=S_{11}.
\end{eqnarray}
and
\begin{eqnarray}
S_{11}&=&S_{22}=\frac{2}{Z_{11}-Z_{12}+Z_{21}-Z_{22}},\nonumber\\
S_{12}&=&-\frac{Z_{11}+Z_{12}+Z_{21}+Z_{22}}{Z_{11}-Z_{12}+Z_{21}-Z_{22}},\nonumber\\
S_{21}&=&\frac{-Z_{11}+Z_{12}+Z_{21}-Z_{22}}{Z_{11}-Z_{12}+Z_{21}-Z_{22}}.
\end{eqnarray}

\section{Quantum wave impedance for a \\ finite cascade of constant potentials}

One of the most important properties of a transfer matrix formalism is that the transfer matrix for the whole region of a piecewise constant potential can be calculated as a product of transfer matrices of each region where a potential has a constant value.
So if a potential has the following form
\begin{eqnarray}
U(x)=
\left\{\begin{array}{cc}
0, & x\le x_0\\
U_i(x),& x_i<x<x_{i+1}, i=1\ldots N \\
0, &  x\ge x_{N+1}
\end{array}\right.
\end{eqnarray}
and we have a transfer matrix $T_i$ for each region $x\in [x_i\ldots x_{i+1}]$.
Then the transfer matrix for a whole region is the product of transfer matrices, where each of them corresponds to the appropriate single region:
\begin{eqnarray}\label{mult_law}
T_{x_0x_{N+1}}=\prod_{j=0}^N T_{x_{j}x_{j+1}}.
\end{eqnarray}
It is the consequence of this simple relation: $T_{x_{i}x_{i+2}}=T_{x_{i}x_{i+1}}T_{x_{i+1}x_{i+2}}$, which we try to explain below. So
the transfer matrix for a single region with a constant potential consists of three multipliers:
\begin{eqnarray}
T_{ab}=I_{12}P_{ab}I_{23}.
\end{eqnarray}
A first matrix $I_{12}$ and a third one $I_{23}$ describe a wave transferring through the interface of regions with different values of a potential energy. The second matrix $P_{ab}$ describes a wave transferring from a point $a$ to a point $b$ inside one region of a constant potential.
Because of very important property of matrices $I_{ij}$ \cite{SanchezSoto_atall:2012}: $I_{ij}=I_{ik}I_{kj}$ we finally get that
\begin{eqnarray}
T_{ac}=I_{12}P_{ab}I_{23}P_{bc}I_{34}=I_{12}P_{ab}I_{21}I_{13}P_{bc}I_{34}=T_{ab}T_{bc}.
\end{eqnarray}

A special case of the multiplication law (\ref{mult_law}) is that when in all $N$ regions the function of a potential energy is the same. Then the transfer matrix of the whole system is the transfer matrix of a single region but of power $N$. Using The Cayley–Hamilton theorem which says that every square matrix satisfies
its own characteristic equation \cite{Griffiths_Steinke:2001}:
\begin{eqnarray}
T^2-2\chi T+I=0,
\end{eqnarray}
where $\chi=\frac{1}{2}Tr(T)$
one can calculate the resulting transfer matrix $T$ in terms of the elements of
the transfer matrix of a single region, which finally gives: 
\begin{eqnarray}
T^N=TU_{N-1}(\chi)-IU_{N-2}(\chi),
\end{eqnarray} 
where $U_N(\chi)$ is the second kind Chebyshev's polynomial of a degree $N$ in $\chi$. This polynomial obeys the recurrent relation:
\begin{eqnarray}
U_{N+2}(\chi)-2\chi U_{N+1}(\chi)
+U_{N}(\chi)=0.
\end{eqnarray}
Notice that a Chebyshev's second kind polynomial can be shown as:
\begin{eqnarray}
U_N(\chi)=\frac{\sin[(N+1)\lambda(\chi)]}{\sin[\lambda(\chi)]},
\end{eqnarray}
where $\lambda=\arccos(\chi)$.
Thus, the transfer matrix of a whole system is
\begin{eqnarray}
T^N=\begin{pmatrix}
T_{11}U_{N-1}(\chi)-U_{N-2}(\chi) \quad T_{12}U_{N-1}(\chi)
\\
T_{21}U_{N-1}(\chi)\quad T_{22}U_{N-1}(\chi)-U_{N-2}(\chi)
\end{pmatrix}.
\end{eqnarray}
It gives the expressions for a reflection and a transmission coefficients
\begin{eqnarray}
r=\frac{T_{21}U_{N-1}(\chi)}
{T_{11}U_{N-1}(\chi)-U_{N-2}(\chi)},\quad
t=\frac{1}
{T_{11}U_{N-1}(\chi)-U_{N-2}(\chi)}
\end{eqnarray}
and the relation between a quantum wave impedance at the beginning of a studied system and amplitudes $A_N$ and $B_N$ of waves at the end of a system. $A_N$ is the amplitude of a wave which propagates in a positive direction of an axis and $B_N$ is the amplitude of a wave which propagates in a negative direction of an axis.
\begin{eqnarray}
Z(0)\!\!\!&=&\!\!\!z_0\left(\frac{}{}[(T_{11}-T_{21})U_{N-1}(\chi)-U_{N-2}(\chi)]A_N-\right.\left.[\frac{}{}(T_{22}-T_{12})U_{N-1}(\chi)-U_{N-2}(\chi)]B_N\right)\times\nonumber\\
\!\!\!&\times&\!\!\!\left(\frac{}{}[(T_{11}+T_{21})U_{N-1}(\chi)-U_{N-2}(\chi)]A_N+\right.\left.[\frac{}{}(T_{22}+T_{12})U_{N-1}(\chi)-U_{N-2}(\chi)]B_N\right)^{-1}\!\!\!.
\end{eqnarray}
Taking into account that 
\begin{eqnarray}
Z(N)=z_N\frac{A_N-B_N}{A_N+B_N}=z_N\frac{1-\rho}{1+\rho},\quad
\rho=\frac{z_N-Z(N)}{z_N+Z(N)}
\end{eqnarray}
we finally get
\begin{eqnarray}
Z(0)&=&z_0\left(\frac{}{}[T_{11}-T_{21}+T_{12}-T_{22}]U_{N-1}(\chi)z_N+\right.\nonumber\\
&+&\left.\frac{}{}(T_{11}-T_{21}+T_{22}-T_{12})U_{N-1}(\chi)-2U_{N-2}(\chi)]Z(N)\right)\times\nonumber\\
&\times&\left(\frac{}{}[(T_{11}+T_{21}+T_{12}+T_{22})U_{N-1}(\chi)-2U_{N-2}(\chi)]z_N-\right.\nonumber\\
&-&\left.\frac{}{}[T_{22}+T_{12}-T_{11}-T_{21}]U_{N-1}(\chi)Z(N)\right)^{-1}.
\end{eqnarray}
We can rewrite this result using elements of a matrix $Z$, which we introduced earlier:
\begin{eqnarray}\label{Z0ZNc}
Z(0)=z_0\frac{\frac{}{}Z_{11}U_{N-1}(\chi)z_N-
	[Z_{12}U_{N-1}(\chi)+U_{N-2}(\chi)]Z(N)}{[Z_{21}U_{N-1}(\chi)-U_{N-2}(\chi)]z_N-\frac{}{}Z_{22}U_{N-1}(\chi)Z(N)}
\end{eqnarray}
keeping in mind that the values $Z_{ij}$ are defined on the base of a relation (\ref{Z_thr_T}).
Thus, the problem of a calculation of a quantum wave impedance for a system of identical regions of a potential is solved. And
once the expression for a quantum wave impedance for a single region is obtained it can be easily extended
to the analytical calculation of a quantum wave impedance for $N$ identical potentials, like it is for a transfer matrix approach.

\section{Finite periodic systems}

The aim of this section is to present an introduction of an application of a quantum wave impedance to finite periodic systems. 

So let's consider the model with the following potential energy:
\begin{eqnarray}
U(x)&=&U_1\theta(-x-l_L-l)+U_0(x)(\theta(x)-\theta(x-L))+U_2\theta(x-L-l_R),
\end{eqnarray}
where $U(x)$ is a periodic function with a period $a$,
$l_L, l_R\geq 0$, 
$L=Nl$, $N$ is the number of elementary cells.

In a previous section we have found the relation (\ref{Z0ZNc}) which in our case can be written as: 
\begin{eqnarray}\label{Z0ZL}
Z(0)=z_0\frac{Z_{11}U_{N-1}(\chi)z_N-[Z_{12}U_{N-1}(\chi)+U_{N-2}(\chi)]Z(L)}{[Z_{21}U_{N-1}(\chi)-U_{N-2}(\chi)]z_N-Z_{22}U_{N-1}(\chi)Z(L)},
\end{eqnarray}
where $Z_{ij}, i,j=1..2$ are determined on the base of the expression which relates the value of a quantum wave impedance at the boundary points of the elementary cell (\ref{Zmatrix_finding}).

Notice that these relations are independent of a specific potential profile, so it can be used to obtain a quantum wave impedance of a periodic system with an arbitrary potential profile including a piecewise constant periodic potential. Having a quantum wave impedance of a periodic system, we are able to get expressions for a transmission and a reflection coefficients.

Then using iterative formulas for a quantum wavimpedance determination we find that
\begin{eqnarray}
Z(-l_L-l)&=&z_0\frac{Z(0)\ch[\gamma_0(l_L+l)]-z_0\sh[\gamma_0(l_L+l)]}
{z_0\ch[\gamma_0(l_L+l)]-Z(0)\sh[\gamma_0(l_L+l)]},\nonumber\\
Z(L)&=&z_N\frac{z_R\ch[\gamma_Nl_R]-z_N\sh[\gamma_Nl_R]}
{z_N\ch[\gamma_Nl_R]-z_R\sh[\gamma_Nl_R]},
\end{eqnarray}
where $\gamma_0=imz_0/\hbar$, $\gamma_N=imz_N/\hbar$.
Expressing $Z(0)$ and $Z(L)$ from these relations, substituting them into (\ref{Z0ZL}) and reminding the condition for eigenenergies determination ($Z(-l_1-l)=-z_L$) we get the following formula:
\begin{eqnarray}\label{Zgen_fps}
&&\frac{z_0\th[\gamma_0(l_L+l)]-z_L}
{z_0-z_L\th[\gamma_0(l_L+l)]}=\left(\frac{}{}Z_{11}U_{N-1}(\chi)(z_N-z_R \th[\gamma_Nl_R])\right.\nonumber\\
&&-[Z_{12}U_{N-1}(\chi)+U_{N-2}(\chi)]
\left.\frac{}{}(z_R-z_N \th[\gamma_Nl_R])\right)\nonumber\\
&&\left(\frac{}{}[Z_{21}U_{N-1}(\chi)-U_{N-2}(\chi)](z_N-z_R \th[\gamma_Nl_R])-\right.\nonumber\\
&&\left.-Z_{22}U_{N-1}(\chi)(z_R-z_N \th[\gamma_Nl_R])\frac{}{}\right)^{-1}.
\end{eqnarray}

\section{Finite Dirac comb}
In this section we discuss a system, properties of which is periodic along the length of the system.
So, consider a finite Dirac comb with a following form of a potential energy:
\begin{eqnarray}
U(x)=U_L\theta(-x-l)+\sum_{n=0}^{N-1} \alpha\delta(x+nl)(\theta(x)-\theta(x-L))+U_R\theta(x-L).
\end{eqnarray}
For the first time this model was solved by Sokolov in 1934 \cite{Sokolow:1934}. He based on the Tamm's approach but extended it and proposed an original way of studying surface states. In this section we will show how to solve this model using a quantum wave impedance approach.

So, for a single cell of a Dirac comb we have
\begin{eqnarray}
Z(a)\!\!\!\!\!&=&\!\!\!\!\!z_0\frac{Z(b)\ch[ik_0l]-z_0\sh[ik_0l]}{z_0\ch[ik_0l]-Z(b)\sh[ik_0l]}+i\frac{2\alpha}{\hbar}=\nonumber\\
\!\!\!\!\!&+&\!\!\!\!\!z_0\frac{z_0(-\sh[-i\xi]+2i\Omega\ch[i\xi])-Z(b)(-\ch[i\xi]+2i\Omega\sh[i\xi])}{z_0\ch[i\xi]-Z(b)\sh[i\xi]},\nonumber\\
\end{eqnarray}
where $\Omega=p/\xi=\alpha/(z_0\hbar)$, $z_0=\sqrt{2mE/2}$, $\xi=k_0l$, $k_0=mz_0/\hbar$,  $p=m\alpha l/\hbar^2$.

On the base of a formula (\ref{Zmatrix_finding}) we have \begin{eqnarray}\label{Zij_Dc}
Z_{11}=-\sh[i\xi]+2i\Omega\ch[i\xi],&&
\quad Z_{12}= -\ch[i\xi]+2i\Omega\sh[i\xi],\nonumber\\
Z_{21}=\ch[i\xi],&&\quad Z_{22}=\sh[i\xi].
\end{eqnarray}

But before proceeding we have to be convinced that $\det|Z|=1$. We can see this by substituting expressions for $Z_{ij}$ (\ref{Zij_Dc}) into the formula for $\det|Z|=Z_{11}Z_{22}-Z_{12}Z_{21}$.

In a case of a considered here system we have that 
$l_L=a$, $l_R=0$, $z_N=z_0$. For a simplification of a process of a calculation let's assume that $z_L=z_R=z=z_E$ ($z_E=\sqrt{2(E-U_E)/m}$). Applying both these and the values for $Z_{ij}$ to the obtained in the previous section formula (\ref{Zgen_fps}) we get the relation for  the determination of eigenenergies, namely 
\begin{eqnarray}
\frac{z_0\th[i\xi]\!-\!z_E}
{z_0\!-\!z_E\th[i\xi]}\!=\!\frac{Z_{11}U_{N-1}(\chi)z_0\!-\![Z_{12}U_{N-1}(\chi)\!+\!U_{N-2}(\chi)]z_E}{[Z_{21}U_{N-1}(\chi)\!-\!U_{N-2}(\chi)]z_0\!-\!Z_{22}U_{N-1}(\chi)z_E}.
\end{eqnarray}
After expressing $U_{N-2}(\chi)/U_{N-1}(\chi)$ we can rewrite the previous relation in the following form:
\begin{eqnarray}
\frac{\!(Z_{21}z_0-Z_{22}z_E)(z_0\th[i\xi]-z_E)-(Z_{11}z_0+Z_{12}z_E)(z_0-z_E\th[i\xi])}{(z_0^2+z_E^2)\th[i\xi]-2z_0z_E}=\frac{U_{N-2}(\chi)}{U_{N-1}(\chi)}.
\end{eqnarray}
Taking into account that
\begin{eqnarray}
U_N(\chi)-2\cos(\chi)U_{N-1}(\chi)+U_{N-2}(\chi)=0
\end{eqnarray}
we find 
\begin{eqnarray}
\frac{U_{N-2}(\chi)}{U_{N-1}(\chi)}&=&2\cos(\chi)-\frac{U_N(\chi)}{U_{N-1}(\chi)}=2\cos(\chi)-
\frac{\sin((N+1)\chi)}{(\sin(N\chi))}=\nonumber\\
&=&2\cos(\chi)-\frac{1}{\cos(\chi)-\sin(\chi)\ctg((N+1)\chi)}.
\end{eqnarray}
And finally we get the following expression for the determination of energies of surface states:
\begin{eqnarray}\label{E_cond_fDC}
&&\!\!\!\!\!\!\!\!\!\!\!\cos(\chi)-\sin(\chi)\ctg((N+1)\chi)=\left[\frac{}{}2\cos(\chi)-\right.\nonumber\\
&&\!\!\!\!\!\!\!\!\!\!\!\!-\!\!\left.\frac{(Z_{21}z_0\!-\!Z_{22}z_E\!)(z_0\!\tg[i\xi]\!-\!z_E\!)\!-\!(Z_{11}z_0\!+\!Z_{12}z_E\!)\!(z_0\!-\!z_E\!\tg[i\xi])}{(z_0^2+z_E^2)\tg[i\xi]-2z_0z_E}\!\right]^{\!-\!1}\!\!\!\!\!,
\end{eqnarray}
where 
\begin{eqnarray}
\cos(\chi)=\cos(\xi)+\Omega\sin(\xi).
\end{eqnarray}
Taking into account the expressions for $Z_{ij}$ (\ref{Zij_Dc}) and that $\cos(\chi)=(Z_{21}-Z_{12})/2$ we get
\begin{eqnarray}
&&\!\cos(\chi)\!-\!\sin(\chi)\ctg((N\!+\!1)\chi)\!=\!
\frac{i(z_0^2\!+\!z_E^2)\sin[\xi]\!-\!2z_0z_E\cos[\xi]}
{2i\Omega z_0^2-2z_0z_E}=\nonumber\\
&&=\frac{(-\varkappa_E^2/k_0^2+1)\sin[\xi]\!-\!2\varkappa_E/k_0\cos[\xi]}
{2(\Omega-k_0/\varkappa_E)}.
\end{eqnarray}

\section{Finite Dirac comb. Transfer matrix app\-roach}

Using relation (\ref{T_thr_Z}) and expressions for reflection amplitudes $r_0$ and $r_N$ \cite{Arx1:2020} at the beginning ($x=-l$) and at the end ($x=L$) of a studied system 
\begin{eqnarray}
r_0=\exp[-2i\xi]\frac{z_0+z_E}{z_0-z_E},\quad
r_N=\frac{z_0-z_E}{z_0+z_E}
\end{eqnarray}
we can rewrite a formula (\ref{E_cond_fDC}) in the following form:
\begin{eqnarray}
&&\cos(\chi)-\sin(\chi)\ctg((N+1)\chi)=\nonumber\\
&&=\frac{r_N-r_0}{r_0T_{11}\!+\!r_0r_NT_{12}\!-\!T_{21}\!-\!T_{22}r_N\!-\!2\cos(\chi)(r_0\!-\!r_N)}.
\end{eqnarray}
Now taking into account that $2\cos(\chi)=T_{11}+T_{22}$ we get
\begin{eqnarray}\label{E_cond_T}
\cos(\chi)\!-\!\sin(\chi)\ctg((N+1)\chi)\!=\!
\frac{r_N-r_0}{r_NT_{11}\!+\!r_0r_NT_{12}\!-\!T_{21}\!-\!T_{22}r_0}.
\end{eqnarray}
Using the relation (\ref{T_thr_Z}) we can calculate the transfer matrix for a single cell of a Dirac comb:
\begin{eqnarray}
T=\begin{pmatrix}
\left(1+i\Omega\right)\exp(-i\xi), & i\Omega\exp(i\xi)
\\
-i\Omega\exp(-i\xi), & 
\left(1-i\Omega\right)\exp(i\xi)
\end{pmatrix}.
\end{eqnarray}
Substituting the elements of the obtained transfer matrix into 
the relation (\ref{E_cond_T}) and after simple transformations we get
\begin{eqnarray}
\sin(\chi)\ctg((N+1)\chi)-\cos(\chi)=\frac{(\varkappa_E^2/k_0^2-1)\sin(\xi)+2\varkappa_E/k_0\cos(\xi)}{2(\beta-\varkappa_E/k_0)}.
\end{eqnarray} 
We also can depict this result in a form used in the paper \cite{Steslicka:1974}
\begin{eqnarray}\label{cotfDC}
\ctg[(N+1)\chi]=\frac{\mu^2-1}{\mu},
\end{eqnarray}
where
\begin{eqnarray}
\mu=\frac{\sin(\chi)}{(\varkappa_E/k_0-\beta)\sin(\xi)},
\end{eqnarray}
and $\varkappa_E=\sqrt{2(U_E-E)/m}$.

\section{Conclusions}
The models of finite periodic systems play very important role in a study of nano\-heterostructures. They give us the possibility to investigate surface states and impurity states analytically. Another widely-investigated problem which is closely related to this topic is the properties of classical waves in such periodically layered structures as phononic and photonic crystals. The character feature of these systems is that the periodic structure can be flexibly designed and shaped, and this fact among other things gives the opportunity to tailor the boundary-dependent states by a specific choice of a surface location.

So, effective methods for a study of finite systems are of great interest. It is obvious that when we move from the one barrier/well system to the double barrier/well system the complexity of a calculation of a quantum wave impedance increases. And naturally one expects the enormous growing a complexity for a big number of barriers/wells. But in this paper we showed that  it is not a case for periodic systems not only within a transfer matrix approach but also within a quantum wave impedance method. We demonstarated how to get compact expressions for a quantum wave impedance of a finite periodic system by using properties and symmetries of a transfer matrix.

Following the aim of finding effective methods of finit periodic systems investigation we derived the relation between a transfer matrix formalize and a quantum wave impedance approach in order to combine the advantages of both methods. The application of obtained results we illustrated on a finite Dirac comb and within developed approach we get the expressions for a  Tamm's levels in this system.

\renewcommand\baselinestretch{1.0}\selectfont


\def\name{\vspace*{-0cm}\LARGE 
	Bibliography\thispagestyle{empty}}
\addcontentsline{toc}{chapter}{Bibliography}

{\small

	\bibliographystyle{gost780u}
	\bibliography{full.bib}
	
}

\newpage

\end{document}